%% file: paper2.tex
\documentstyle[12pt]{article}
\begin{document}
\baselineskip0.8cm
\input{psbox.tex}

\catcode`\æ=\active
\def æ{{\alpha}}
\catcode`\ß=\active
\def ß{{\beta}}
\catcode`\Þ=\active
\def Þ{{\gamma}}
\catcode`\=\active
\def {{\delta}}
\catcode`\Ê=\active
\def Ê{{\epsilon}}
\catcode`\Â=\active
\def Â{{\zeta}}
\catcode`\Ì=\active
\def Ì{{\eta}}
\catcode`\ý=\active
\def ý{{\theta}}
\catcode`\È=\active
\def È{{\kappa}}
\catcode`\¬=\active
\def ¬{{\lambda}}
\catcode`\µ=\active
\def µ{{\mu}}
\catcode`\×=\active
\def ×{{\nu}}
\catcode`\‰=\active
\def ‰{{\xi}}
\catcode`\¼=\active
\def ¼{{\pi}}
\catcode`\®=\active
\def ®{{\rho}}
\catcode`\Ò=\active
\def Ò{{\sigma}}
\catcode`\Ý=\active
\def Ý{{\tau}}
\catcode`\º=\active
\def º{{\upsilon}}
\catcode`\©=\active
\def ©{{\xi}}
\catcode`\Ÿ=\active
\def Ÿ{{\varphi}}
\catcode`\Ï=\active
\def Ï{{\phi}}
\catcode`\Ú=\active
\def Ú{{\psi}}
\catcode`\Ð=\active
\def Ð{{\omega}}
\catcode`\þ=\active
\def þ{{\Gamma}}
\catcode`\=\active
\def {{\Delta}}
\catcode`\‡=\active
\def ‡{{\Theta}}
\catcode`\'=\active
\def '{{\Theta}}
\catcode`\|=\active
\def |{{\Lambda}}
\catcode`\Ž=\active
\def Ž{{\Xi}}
\catcode`\½=\active
\def ½{{\Pi}}
\catcode`\…=\active
\def …{{\Sigma}}
\catcode`\·=\active
\def ·{{\Phi}}
\catcode`\=\active
\def {{\Psi}}
\catcode`\ð=\active
\def ð{{\Omega}}
\catcode`\«=\active
\def «{{$^7 Be$}}
\catcode`\#=\active
\def #{{\Large{\P}}}
\catcode`\@=\active
\def @{{tel.}}
\catcode`\è=\active
\def è{{\`e}}

\input{title.tex}
\section{Introduction.}

The group led by F.Vannucci has recently published the results of a search for
radiative decays $\nu_1 \to \nu_2 Þ$, where $×_1$ is (in principle)
any neutrino present in the solar flux, either directly produced or arising
from oscillations. To suppress obvious background, the experiment was conducted
during a total solar eclipse and was in practice sensitive to radiative
decays taking place between the Moon and the Earth. Their analysis of the data
was conducted in the specific case where $×_1$ and $×_2$ are nearly degenerate (
$m^2$ $\equiv m_{\nu_1}^2 - m_{\nu_2}^2 \simeq 10^{-5} \,{\rm eV^2 \,
_{/c^4}}$).

In the present note, we present the kinematics for the general case, and work
out in details the limit $m_{\nu_2} \simeq 0$.

We interpret the decay in terms of a dipole (magnetic or electric)
transition, taking into account the polarization of the neutrinos. Results are
then confronted to earth-based limits on transition magnetic moments. The current
sensitivity (assuming only one photon in the decay) falls way below current
limits if either $×_1$ or $×_2$ is one of the known neutrinos, but the bounds
apply here independently of the neutrino species. Possible improvements of the
experiment involve  satellite-based observations.

In section 2 and 3, we discuss the geometry and the kinematics leading to the
observation of a visible photon, as a function, notably, of the mass
difference and the telescope aperture. We discuss the possible signal
configurations, in particular for the monochromatic $×_1$ from «, using standard
solar model data. We consider also the angular distribution of neutrino decays
for the case of polarized neutrinos, and we relate the observed $Þ$ flux to the
transition magnetic moment.

In section 4 to 6, we discuss the relation to existing bounds from earth-based
experiments, and present some prospects for a satellite-bound extension of the
experimental set-up.

\section{Kinematics.}

Let us consider the effective lagrangian (Pauli term) describing the
reaction $\nu_1 \to \nu_2 Þ$:
\begin{eqnarray}
{\cal L} = µ \; \bar Ú_{\nu_2}\; Ò^{æ ß} Ú_{\nu_1} \;
 q_{ß}\;A_æ
\end{eqnarray}
\noindent where $µ$ is the transition magnetic moment $µ_{\nu_1 \nu_2}$ and q
the momentum of the photon.

The correspondent differential decay probability in the neutrino rest
frame (CM) reads:
\begin{eqnarray}
d þ ^* =  {µ^2  \over (2 ¼)^{2}} {^4 (k_1 - k_2 -
k_Þ) \over 2 Ð_1 2 Ð_2 2 Ð_Þ} (8 m_1 k_Þ^2
(Ð_2 + k_2) A(ý^*)) \; d^3 k_2 d^3 k_Þ 
\end{eqnarray}
\noindent where $k_1,k_2,k_Þ$ are the 4-momenta of $×_1$,$×_2$,$Þ$ respectively;

$\,\,\,\,\,\,\,\,\,Ð_1,Ð_2,Ð_Þ$ , their energies;

\noindent and $\,\,\,\,\,\,ý^*$ , the angle between the photon and the boost
direction (defined in the CM as the direction opposite to the sun recoil).

$A(ý^*)$ is found to be:\begin{tabbing}
aaaaaaaaaaaaaaaaaaaaaaa \= \kill
\hspace{-0.8cm}
\> 1 for an unpolarized neutrino flux; \\
\> $1-\cos ý^*$ for a left-handed neutrino flux; \\
\> $1+\cos ý^*$ for a right-handed neutrino flux.
\end{tabbing}

In the rest frame, we have:
\begin{eqnarray}
{d þ^* \over d ð^*} = {µ^2 \over (2 ¼)^2} A(ý^*) \left({
m^2 \over 2 m_1} \right)^3  
\end{eqnarray}
\noindent where $ð^*$ is the solid angle of the photon in the rest frame.

A last integration gives us the decay probability and the mean lifetime:
\begin{eqnarray}
þ^* &=& {µ^2 \over  ¼} \left({ m^2 \over 2m_1} \right)^3 \nonumber \\
Ý_{×_1} &=& K \cdot \left({ m^2 \over {\rm eV^2} }\right) ^{-3} \cdot \left({
m_1 \over {\rm 1 eV}} \right)^3 \cdot \left({µ \over 10^{-10} µ_B }\right) ^{-2}
\end{eqnarray}
\noindent where $K$ is given by ${8 \over æ}\,m_e^2\,\hbar \simeq 1.00 \times
10^{15}$s. {\cite{tau}}

\section{Geometry.}

We calculate first the image produced by a point-like source (taking advantage
of the decay symmetry) at C, the center of the Sun. A point-like source located
in Earth sky at angles ($ý , Ÿ$) produces a translated image, centered at
($ý , Ÿ$ ) instead of ($0,0$), and the full image results of the sum of the
contributions from all pointlike sources inside the Sun.

The source S emits a neutrino which eventually oscillates (but holds it initial
flight direction) and decays at P. There, it emits a photon with an angle $ý$
with respect to its initial flight direction. This photon is collected
in the telescope T on earth. This simplified geometry (source and telescope are
assumed point-like) is plane (cylindrical symmetry).

Four elements will help to solve our problem.

\begin{itemize}
\item[1.]The energy $E$ of the solar neutrinos is typically of order 1 MeV. With
a typical neutrino mass $m_1$ of order 1 eV, we get the following Lorentz
parameters, connect the CM and the laboratory (LAB) frames:{\mbox {$ß
=1-\epsilon\,$}},{\mbox {$\, Þ = {E \over m_1} \simeq 10^6$}}, with $\epsilon
\simeq 10^{-12}$.

Therefore, if $k,k^*$ denote the energies of the photon respectively in the LAB
and the CM,
\begin{eqnarray}
\label{k}
k = k^* Þ (1+ß \cos ý^*) \simeq k^* Þ (1 + \cos ý^*) 
\end{eqnarray}
\noindent valid as long as the Lorentz parameter $ß$ is close to 1.

\item[2.] Eq.(\ref{k}) shows that the energy of the emitted photon
is related to the emission angle $ý$ in the LAB by:
\begin{eqnarray}
\label{theta}
ý \simeq \tan ý &\simeq & {1 \over Þ} \sqrt{{2 k^* Þ \over
k} - 1}\nonumber \\
& \simeq & { m \over E} \cdot \sqrt{ { m^2 E \over
k\; m^2} - 1}\nonumber \\  
& \simeq & { m \over E } \cdot \sqrt { { k_{max} \over k } -1 }
\end{eqnarray}
\noindent where $k_{max} \equiv 2 k^* Þ$ is the maximum photon energy
reachable in the LAB.

\item[3.] The set of points P with S,T and $ý$ (thus $k$) fixed is an arc of a
circle. Since we consider only decays between the Moon and the
Earth\footnote{It is in principle conceivable that neutrinos further than the
Moon, but off-center, contribute by their decay. In practice however, the
limited aperture of the telescope used in \cite{van} forbids their
observation.}, this arc may be treated in a good approximation as a straight
line. Moreover, this line is tangent to the arc at T and makes with the
Sun-Earth direction an angle $ý$ of the same amplitude than the emission angle.

\item[4.] The approximation above, valid as long as $d_{ST} \gg d_{PT}$, also
implies that the results don't depend on the distance of the source S, so that
only its angular position, as seen on Earth sky is relevant. In other terms,
all neutrino sources located in a small rod of the sun, defined by the angles
$(ý,Ÿ )$ on earth, give superimposed images. From now on, such a rod will be
called ''rod-source`` and it image, $Image(ý,Ÿ)$.
\end{itemize}

If the telescope of angular aperture $ý_@$ (corresponding to an energy
$k_@$), is sensitive in a given energy spectrum, for example [ $k_{red}$ ;
$k_{violet}$ ] for the visible spectrum, only the photons between two angles
$ý_{k_+}$ and $ý_{k_-}$ will be detected, where $k_- = max(
k_{red} ; k_@)$ and $k_+ = min(k_{violet} ; k_{max})$. We
see that a rod-source will produce a crown on the telescope (centered on (0,0)
for a source at C). That crown may be so small that it degenerates into a
single pixel (``pixel case'') or can be larger than the telescope aperture
(``out case'').
$$
\hspace{-2.5cm}
\boxit{0pt}{\psboxto(12cm;0cm){paper01.eps}} $$
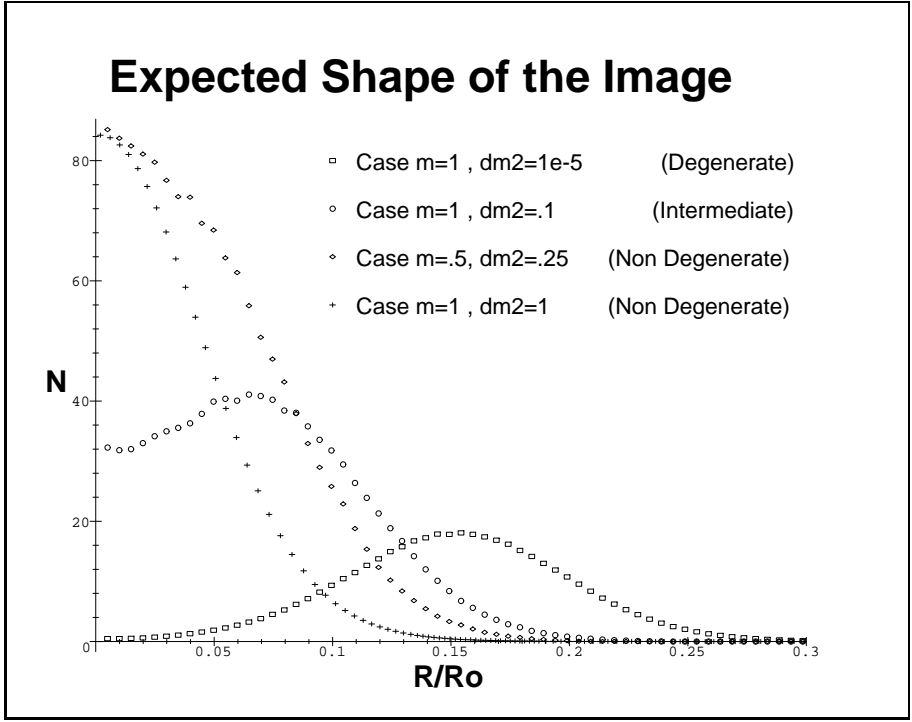
{\small \makebox[1cm]{ } \ \parbox{10cm}{Figure 1 : Expected image shapes for
various values of m$_{×_1}$ (m, in eV) and $m^2$ (dm2, in eV$^2$). N is in
arbitrary units.}}

\bigskip

To restore the full geometry, we have to sum the different contributions
from rod-sources. The data of the standard solar model (SSM)
list usually the production rates as a function of the distance $R$ from the
sun's center. We have to reparametrize the data to
work out the total emission in a rod-source,i.e. a thin
cylinder in the Sun, approximately parallel to the Sun-Earth direction (errors
less than 0.1 $\%$).
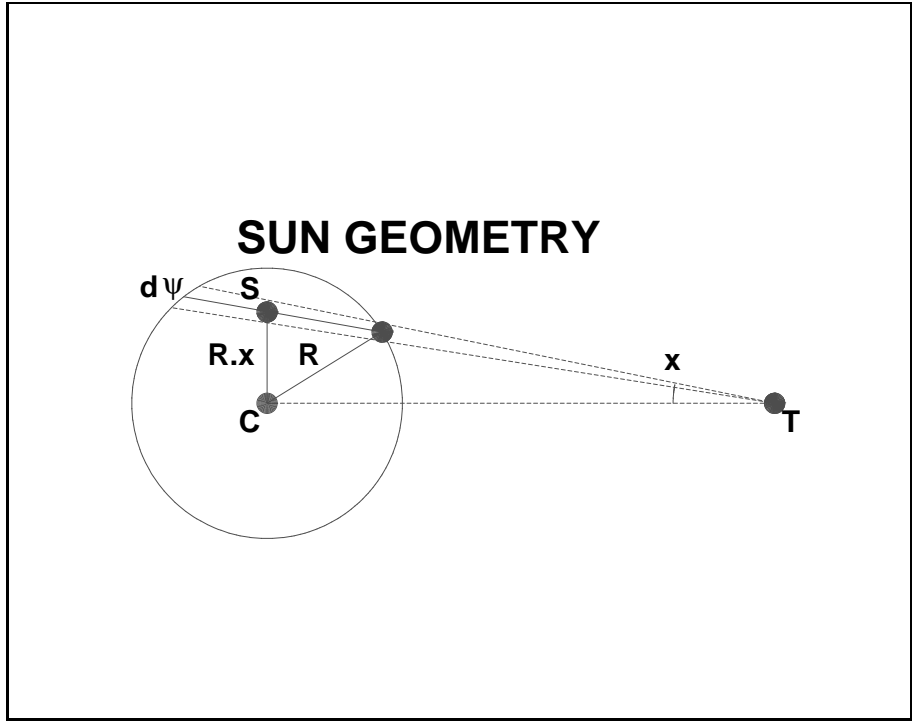
$$
\hspace{-.5cm}
\boxit{0pt}{\psboxto(12cm;0cm){ps01.eps}}
$$
Let $f(R)$ be the solar neutrino flux produced per unit volume at a distance R
from the centre of the Sun ($f(R) = {dÏ (R) \over 4 ¼ R^2 dR}$ where
$dÏ(R)$ are the usual form of the SSM data, see {\cite{bah}}). The emission in
a thin rod in a solid angle $dÚ$ oriented in a direction ($ý,Ÿ$) from
Earth sky is independent of $Ÿ$ and given by\footnote{$x$ is a reduced coordinate
($R \over R_\odot$ or $ý \over ý_\odot$). Here $x$ means the reduced distance
between the rod and C, while $x^\prime$ in (\ref{ec},\ref{ic}) is the distance
between the source S and C.}:
\begin{eqnarray} \label{ec}
 Ê(x) = 2 \;\left(\frac{d_{ST}}{R_\odot} \right) ^2 \; d Ú \int_{x} ^{1}
f(x^\prime) {x^\prime d x^\prime \over \sqrt{x^{\prime2} - x^2}}
\end{eqnarray}
\noindent where $R_\odot$ is the radius of the Sun and $d_{ST}$ is the distance
between the source S and the telescope T (approx. distance Sun-Earth).

We sum now the contributions from all solar sources
\begin{eqnarray}
\label{ic}
 2 \;\left(\frac{d_{ST}}{R_\odot}\right)^2 \int_0 ^{ý_\odot} \sin ý \;d ý \int_0
^{2¼} d Ÿ  \, Image(ý,Ÿ) \cdot \int_{x} ^{1} f(x^\prime) {x^\prime
dx^\prime \over \sqrt{x^{\prime 2} - x^2}}
\end{eqnarray}
We also calculate the expected number of photons landing in the
telescope (assuming an ideal efficiency of 100\%).
\begin{eqnarray}
N_Þ ={ T_{exp} \over \hbar }  \cdot {F \over ß c}\cdot \int_{ _{\rm
{\stackrel{ emitting}{ _{volume}}}}} \!\!\!\!\!\!\!\!\!\!\!d v \cdot {dþ \over d ð } \cdot { ð} 
\end{eqnarray}
\noindent where 
\begin{description}
\item $T_{exp}$ is the exposition time, and we have restored the dimensional
factor $\hbar$.
\item $\frac{F}{ß c}$ is the number of neutrinos present in a unit volume
between Moon and Earth.
\item ${dþ \over d ð } \cdot { ð}$ is the
probability of a decay detection. $ ð$ is the solid
angle under which the telescope is seen from the decay point P; it's value is $¼
\frac{r_@^2}{d_{PT}^2}$ and varies in space.
\end{description}

The coordinates $ý$ in $dv = r^2drdŸ
d \cos ý$ and in $ ð$ are in principle different. The first one locates P from
Earth sky. The second one locates the photon from P. But we already pointed out
that those angle have the same value.

Substituting $\int dv$ by $\int r^2 dr \int dð$ and $dþ \over
dð$ by $\frac{1}{Þ} \frac{dþ^*}{dð^*}
\frac{dð^*}{dð}$, we obtain
\begin{eqnarray}
\label{N0}
N_Þ ={T_{exp} \over \hbar }\cdot {F \over ß c} \cdot {m \over E}\cdot 
{µ^2 \over {4¼}} { \left({ m^2 \over 2 m} \right)^3 }\cdot
\int_0 ^{2¼}\!\! d Ÿ \cdot \int_A ^B \!\!d \cos{ ý^*
}\,A(ý^*)\,\int_0 ^{d_{ME}}\!\!{r^2 dr } \cdot{ r_@^2\over r^2}
\end{eqnarray}
\noindent where $d_{ME}$ is the distance between the Moon and the Earth.

The first integral gives us a factor $2 ¼$; the third one, $r_@^2 \cdot d_{EM}$;
the second one depend on $A(ý^*)$:it gives ${2  k \over k_{max}} \cdot H$.
Putting into (\ref{N0}): 
\begin{eqnarray}
\label{N}
 N_Þ = {1 \over 8\hbar c}\cdot \left( t \;  k\;r_@^2\;
d_{ME} \right) \cdot \left({F \over E^2}\right) \cdot \left(( m^2)^2 \;
µ^2 \cdot H\right)
\end{eqnarray}
\noindent where $H$ is, for unpolarized, left-handed  and right-handed
neutrinos respectively, $2  k \over k_{max}$, ${2  k \over k_{max}}\cdot \left(
2-\frac{k_- + k_+}{k_{max}}\right) $, and ${2  k \over k_{max}}\cdot \left(
\frac{k_- + k_+}{k_{max}}\right)$. The first bracket gathers the experimental
parameters; the second, the SSM; and the third describes the neutrino physics.

Note the independence on $m$, and the somewhat unexpected dependence on $E$:
the lower energies are favoured\footnote{This has some unexpected
consequences. For example, in the $pp$ neutrinos, however the flux at low
energy (a few keV) is very small, we expect a more intensive emission at low
energy than at high ($> 100 keV$)}. A limit on $ m^2 \cdot µ$ is obtained
from the oberved $N_Þ$ and at fixed $N_Þ$, the limit on $µ$ varies like the
inverse of $m^2$.

It is easy to generalize all this results to a spectrum of neutrino energies. We
only need to substitute $F$ by $\int dE f(E)$ where $f(E)$ is the flux per energy
unit.

\section{Results.}

An analysis of experimental data should involve two aspects: the shape of the
expected signal and the number of photons collected.

As already discussed, the shape of the signal is a centered crown, which may
however be concentrated into a single pixel, or, as another extreme case, fall
out of the telescope aperture. A full analysis must also consider the entirety
of the neutrino spectrum (and not only those produced by the «).
The informations obtained from the shape above concern especially $m^2$. Indeed
$µ$ doesn't appear in (\ref{theta}), and $m$ is a less sensitive parameter.

In the ``pixel'' or ``out'' cases, only limits can be obtained. The pixel case
provides also informations about the SSM, because the shape of the signal
reproduces $e(x)$.

The number of photons provides informations on the quantity $m^2 \cdot µ$. For a
fixed value of $m^2$, one can extract $µ_{×_1×_2}$ and thus $Ý_{×_1}$.

For example, for a ${\mbox{\sc \o}}1m$ telescope and an exposition of 1s., assuming that
the optics allow for collecting of all visible photons falling on the
telescope mirror, considering the full « neutrino flux alone (all neutrinos
assumed to be $×_1$), and an unpolarized neutrino with $m=$ 1eV, the limit
$N_Þ<1000$ provides:  %
 \begin{eqnarray}
m^2 \cdot µ &<&  1.44 \;{\rm eV}^2\; µ_B \nonumber\\
{\rm For }\;m^2 \simeq 1 \;\;\;\;\;\;\;\;\;\;\;\; µ_{×_1×_2} &<& 1.44 \;µ_B    
\nonumber\\ Ý_{×_1} &>& 0.482 \times 10^{-5\;}s.    \nonumber\\
{\rm For }\;m^2 \simeq 10^{-5} \;\;\;\;\;\;\; µ_{×_1×_2} &<& 1.44 \times
10^5\; µ_B     \nonumber\\
 Ý_{×_1} &>& 0.482  \;s.
\end{eqnarray}
We see that the nearly degenerate masses give a better limit on $Ý_{×_1}$. The non
degenerate masses offers less acceptance because the major part of the photons
are now $Þ$-rays; but the decay probability depends on $(m^2)^3$. That case
provides thus a better limit on $µ_{×_1×_2}$.

The signal will fall out of the telescope aperture if $ý_@ <
ý_{violet}$, i.e. if $m^2 > 17$ eV$^2$ (for an aperture of $.5 \times
10^{-2} $rad.). The photon emission will be entirely in the infrared region if
$k_{max} < k_{red}$, i.e. if $m^2 < 2.1 \times 10^{-6} $ eV$^2$.

Note such an experiment covers a large domain of $m^2$ because $E$ varies
from a few keV to about 10 MeV. For example, it is possible that the image from
the « falls out of the telescope but that the $pp$ neutrinos produce a crown.

The detection of unpolarized photons at a given wavelength is not sensitive to
the polarization of the incident neutrinos. To do this, the best way is to
accumulate data at different wavelengths probing the $k$ dependence of $H$ in
(\ref{N}).

\section{Comparison to earth-bound limits.}
If indeed the radiative decay of sun-born neutrinos proceeds through a single
photon emission, the limits on lifetimes are equivalent to limits on
magnetic transition dipole moments (or a combination of electric and magnetic
moments). As discussed in \cite{vys} strong limits on those transition moments
exist, and can be derived directly from the searches for "diagonal" magnetic
moments. These limits are for instance, for any decay involving an electronic
left-handed neutrino $\mu_{i\, e_L} \leq 1.8 \times 10^{-10}\,  \mu_B$, and are
out of reach at least for eclipse-linked experiments. This would be the case,
either for the electron neutrinos originating from the sun and decaying into
$\nu_i$, or for the neutrino of type $i$, ($i \neq e, \mu,\tau$) which would
have already been created through oscillations, decaying to $×_{e_L}$.

The limits obtained from eclipse or space-based experiments however cover the
extra case of an hypothetical $\nu_i$ decaying into $\nu_j$, where neither $i$
nor $j$ refer to left-handed known neutrinos (for example, a new $\nu_i$ decaying
to a right-handed  $\nu_e$).

Reference \cite{vys} also raised the point of the momentum dependence of the
magnetic transition form factor; indeed, accelerator or even reactor experiments
measure $\mu$ at very large momentum transfer with respect to the real decay
processes considered here (photon on shell), and the extrapolation may not be
trivial; we will however not elaborate on this possibility here, since it was
shown that extraordinary assumptions would be needed to bring any serious
difference.

\section{Prospects.}

Earth-based experiments can obviously be improved through larger telescopes,
airplane-borne equipment, but satellite-based experiments offer more
possibilities: $X$ and $Þ$-rays detection; longer and recurrent expositions;
strong reduction of noise;...

For example, an earth-based experiment can involve more than one telescope to
increase the exposition time. If the eclipse occults an observation centre,
bigger mirrors can be used. The optical efficiency and the CCD resolution can
be increased. But on Earth, the rarity of total solar eclipses and the important
atmospheric noise are intrinsic constraints.

On the contrary, satellite-based experiment provides the advantage of recurrent 
data taking, because the Earth occults the satellite every day. Moreover, the
allowed aperture of the telescope will be increased. For a geostationary orbit
(at 36 000 km), the decay volume is reduced by a factor $\sim 10$ but the
satellite is occulted during about half an hour per day and an aperture of
$0.15$ rad. is allowed. Further gain comes from the atmospheric noise
reduction and the wide accessible spectrum, including $X$, $Þ$ and $IR$.

We thank M.Arnould, P.Castoldi, F.Vannucci and P.Vilain for their helpful
discussions and supports.

\end{document}

%% file: psbox.tex
%
%
\catcode`\@=11 
%
%
\def\psfortextures{
\def\PSspeci@l##1##2{%
\special{illustration ##1 scaled ##2}%
}}
\def\psfordvitops{
\def\PSspeci@l##1##2{%
\special{dvitops: import ##1 \the\drawingwd \the\drawinght}%
}}
\def\psfordvips{
\def\PSspeci@l##1##2{%
\d@my=0.1bp \d@mx=\drawingwd \divide\d@mx by\d@my%
\includegraphics{##1}%
}}
\def\putsp@ce#1{#1 }
\def\psforoztex{
\def\PSspeci@l##1##2{%
\special{##1
      ##2 1000 div dup scale
      \putsp@ce{\number-\psllx} \putsp@ce{\number-\pslly} translate
}%
}}
\psfordvips
%
\newread\psiz@
\newdimen\drawinght\newdimen\drawingwd
\newdimen\psxoffset\newdimen\psyoffset
\newbox\drawingBox
\newif\ifNotB@undingBox
\newhelp\PShelp{Proceed: you'll have a 5cm square blank box instead of 
your drawing (Jean Orloff).}
\def\s@tsize#1 #2 #3 #4\@ndsize{
  \drawinght=#4bp\advance\drawinght by-#2bp
  \drawingwd=#3bp\advance\drawingwd by-#1bp
  \def\psllx{#1}\def\pslly{#2}%
  \def\psurx{#3}\def\psury{#4}
  }
\def\sc@nline#1:#2\@ndline{\edef\p@rameter{#1}\edef\v@lue{#2}}
\def\g@bblefirstblank#1#2:{\ifx#1 \else#1\fi#2}
\def\psm@keother#1{\catcode`#112\relax}
\def\execute#1{#1}
{\catcode`\%=12
\xdef\B@undingBox{
}  		
\def\ReadPSize#1{
 \edef\PSfilename{#1 }
 \openin\psiz@=#1\relax
 \ifeof\psiz@ \errhelp=\PShelp
   \errmessage{I haven't found your postscript file (\PSfilename):
if you don't put it next to your input file, you should know how to
redirect TeX's \read command through a jungle of directories...}
   \closein\psiz@
   \drawinght=5cm \drawingwd=5cm   
 \else
   \loop
     \execute{\begingroup
       \let\do\psm@keother
       \dospecials
       \catcode`\ =10
       \catcode`\^^M=9
       \global\read\psiz@ to\n@xtline
       \endgroup}
     \ifeof\psiz@
       \errhelp=\PShelp
       \errmessage{(\PSfilename) is not an Encapsulated PostScript File:
           I could not find any \B@undingBox: line.}
       \edef\value{0 0 142 142:} \let\PSfilename=\null
       \NotB@undingBoxfalse
     \else
       \expandafter\sc@nline\n@xtline:\@ndline
       \ifx\p@rameter\B@undingBox\NotB@undingBoxfalse
       \else\NotB@undingBoxtrue
       \fi
     \fi
   \ifNotB@undingBox\repeat
   \closein\psiz@
   \edef\int@rmediateresult{\expandafter\g@bblefirstblank\v@lue}
   \expandafter\s@tsize\int@rmediateresult\@ndsize
 \fi
\message{#1}
}
%
%
\newcount\xscale \newcount\yscale
\newdimen\d@mx \newdimen\d@my
\def\psboxto(#1;#2)#3{\vbox{
   \ReadPSize{#3}
   \divide\drawingwd by 1000
   \divide\drawinght by 1000
   \d@mx=#1
   \ifdim\d@mx=0pt\xscale=1000
         \else \xscale=\d@mx \divide \xscale by \drawingwd\fi
   \d@my=#2
   \ifdim\d@my=0pt\yscale=1000
         \else \yscale=\d@my \divide \yscale by \drawinght\fi
   \ifnum\yscale=1000
         \else\ifnum\xscale=1000\xscale=\yscale
                    \else\ifnum\yscale<\xscale\xscale=\yscale\fi
              \fi
   \fi
   \divide \psxoffset by 1000\multiply\psxoffset by \xscale
   \divide \psyoffset by 1000\multiply\psyoffset by \xscale
   \multiply\drawingwd by\xscale \multiply\drawinght by\xscale
   \ifdim\d@mx=0pt\d@mx=\drawingwd\fi
   \ifdim\d@my=0pt\d@my=\drawinght\fi
   \message{scaled \the\xscale}
 \hbox to\d@mx{\hss\vbox to\d@my{\vss
   \setbox\drawingBox=\hbox to 0pt{\kern\psxoffset\vbox to 0pt{
      \kern-\psyoffset
      \PSspeci@l{\PSfilename}{\the\xscale}
      \vss}\hss}
   \ht\drawingBox=\the\drawinght \wd\drawingBox=\the\drawingwd
   \copy\drawingBox
 \vss}\hss}
  \global\psxoffset=0pt
  \global\psyoffset=0pt
}}  
%
%
\def\psboxscaled#1#2{\vbox{
  \ReadPSize{#2}
  \xscale=#1
  \message{scaled \the\xscale}
  \divide\drawingwd by 1000\multiply\drawingwd by\xscale
  \divide\drawinght by 1000\multiply\drawinght by\xscale  
  \divide \psxoffset by 1000\multiply\psxoffset by \xscale
  \divide \psyoffset by 1000\multiply\psyoffset by \xscale
  \setbox\drawingBox=\hbox to 0pt{\kern\psxoffset\vbox to 0pt{
     \kern-\psyoffset
     \PSspeci@l{\PSfilename}{\the\xscale}
     \vss}\hss}
  \ht\drawingBox=\the\drawinght \wd\drawingBox=\the\drawingwd
  \copy\drawingBox
  \global\psxoffset=0pt
  \global\psyoffset=0pt
}}  
%
\def\psbox#1{\psboxscaled{1000}{#1}}
%
%
\def\centinsert#1{\midinsert\line{\hss#1\hss}\endinsert}
\def\pscaption#1#2#3{\vbox{
   \setbox\drawingBox=#1
   \hbox{#2\copy\drawingBox}
   \vskip2em
   \vbox{\hsize=\wd\drawingBox\setbox0=\hbox{#3}
     \ifdim\wd0>\hsize
       \noindent\unhbox0
    \else\centerline{\box0}
    \fi
}}}
\def\psfigurebox#1{\pscaption{\psbox{#1}}}
%
\def\at(#1;#2)#3{\rlap{\kern#1\raise#2\hbox{#3}}}
%
\newdimen\gridht \newdimen\gridwd
\def\gridfill(#1;#2){
  \setbox0=\hbox
  to 1truecm{\vrule height1truecm width.4pt\leaders\hrule\hfill}
  \gridht=#1
  \divide\gridht by \ht0
  \multiply\gridht by \ht0
  \gridwd=#2
  \divide\gridwd by \wd0
  \multiply\gridwd by \wd0
  \advance \gridwd by \wd0
  \vbox to \gridht{\leaders\hbox to\gridwd{\leaders\box0\hfill}\vfill}}
%
\def\fillinggrid{\at(0cm;0cm){\vbox{
  \gridfill(\ht\drawingBox;\wd\drawingBox)}}}
%
%
\def\textleftof#1:{
  \setbox1=#1
  \setbox0=\vbox\bgroup
    \advance\hsize by -\wd1 \advance\hsize by -2em}
\def\textrightof#1:{
  \setbox0=#1
  \setbox1=\vbox\bgroup
    \advance\hsize by -\wd0 \advance\hsize by -2em}
\def\endtext{
  \egroup
  \hbox to \hsize{\valign{\vfil##\vfil\cr%
\box0\cr%
\noalign{\hss}\box1\cr}}}
%
\def\boxit#1#2{\hbox{\vrule\vbox{
  \hrule\vskip#1\hbox{\hskip#1\vbox{#2}\hskip#1}%
        \vskip#1\hrule}\vrule}}
\catcode`\@=12 

%% file: title.tex
\begin{titlepage}
\begin{center}
{\Large \bf Probing radiative solar neutrinos decays.}\\
\vspace*{0.6cm}
J.-M Fr\`ere, D. Monderen.\\
{\footnotesize\it Service de Physique Th\' eorique, Universit\' e Libre de
Bruxelles.}\footnote{postal address:Physique Th\' eorique ULB CP 225, Boulevard
du Triomphe, B-1050 Bruxelles, Belgium}\end{center}
\vspace*{1cm} \begin{abstract}
Motivated by a pilot experiment conducted by F.Vannucci et al. during a solar
eclipse, we work out the geometry governing the radiative decays
of solar neutrinos. Surprisingly, although a smaller proportion of the photons
can be detected, the case of strongly non-degenerate neutrinos brings better
limits in terms of the fundamental couplings. We advocate satellite-based
experiments to improve the sensitivity.\end{abstract} \end{titlepage}